\documentclass[%
12pt,
tightenlines,
eqsecnum,
float,
nofootinbib,
amsmath,amssymb,
aps,
prd,
shownopacs
floatfix,
]{revtex4}

\usepackage{graphicx}
\usepackage{dcolumn}
\usepackage{bm}

\usepackage{array}
\usepackage{booktabs}
\usepackage{tabu}
\usepackage{dcolumn}
\usepackage{amsmath}
\usepackage{amsfonts}
\usepackage{amssymb}
\usepackage{graphicx}
\usepackage{subfigure}
\usepackage{graphicx}
\usepackage{dcolumn}
\usepackage{bm}
\usepackage{xcolor}

\graphicspath{{figs/}} 

\begin{document}


\title{Exploring the pre-inflationary dynamics in loop quantum cosmology with a DBI scalar field}

\author{Abolhassan Mohammadi}
  \affiliation{Institute for Theoretical Physics and Cosmology, College of Science, Zhejiang University of Technology, Hangzhou 310032, China.}
 \email{abolhassanm@zjut.edu.cn; abolhassanm@gmail.com}

\date{\today}

\begin{abstract}
Loop quantum cosmology is a symmetry-reduced application of loop quantum gravity. The theory predicts a bounce for the universe at the Planck scale and resolves the singularity of standard cosmology. The dynamics is also governed by an effective Hamiltonian, which predicts a modified Friedmann equation containing the quadratic terms of the energy density. The term plays an essential role in the high energy regime, but the equations return to the standard form in the low energy regime. The evolution of the universe in the pre-inflationary period is studied in the framework of loop quantum cosmology, where the DBI scalar field is assumed to be the dominant component of the universe. Using the numerical method, we provide the evolution of the DBI field. The background evolution shows that there are three phases as: bouncing phase, transition phase and slow-roll inflationary phase. There is also a short period of super-inflation just at the beginning of the bounce phase. The field first climbs the potential and then reaches the turning point where $\dot{\phi}$ disappears and the potential energy becomes the dominant part of the energy density. This is the time when the slow roll inflation begins and the field slowly rolls down the potential. The results indicate that there are a few e-fold expansions in the bounce phase, about $N = 3.5-4$, and the universe experiences about $N = 59$ e-fold expansions in the slow-roll inflation phase.
\end{abstract}

\maketitle


\section{Introduction}\label{sec:introduction}
The inflationary scenario is known as one of the well-known scenarios of the early universe, which has received a huge amount of observational support in the last decades \cite{Planck:2013jfk,Ade:2015lrj,Akrami:2018odb}. The scenario is based on general relativity, and since its first proposal and revision \cite{starobinsky1980new,Guth:1980zm,albrecht1982cosmology,linde1982new,linde1983chaotic}, it has been modified and developed in different ways and in different gravity models \cite{Barenboim:2007ii,Franche:2010yj,Unnikrishnan:2012zu,Rezazadeh:2014fwa,Saaidi:2015kaa,
Fairbairn:2002yp,Mukohyama:2002cn,Feinstein:2002aj,Padmanabhan:2002cp,Aghamohammadi:2014aca,
Spalinski:2007dv,Bessada:2009pe,Weller:2011ey,Nozari:2013wua,Nazavari:2016yaa,
maeda2013stability,abolhasani2014primordial,alexander2015dynamics,Tirandari:2018xyz,
maartens2000chaotic,golanbari2014brane,Mohammadi:2020ake,Mohammadi:2020ctd,Mohammadi:2015upa,
Mohammadi:2015jka,
berera1995warm,berera2000warm,hall2004scalar,Sayar:2017pam,Akhtari:2017mxc,Sheikhahmadi:2019gzs,Rasheed:2020syk,
Mohammadi:2018oku,Mohammadi:2019dpu,Mohammadi:2018zkf,Mohammadi:2019qeu,Mohammadi:2020ftb,
Mohammadi:2021wde,Mohammadi:2021gvf,Mohammadi:2022vru,Mohammadi:2022fiv}. According to the theory, the scalar field, which is assumed to be the dominant element of the universe at the time, will drive the universe through an extreme phase of accelerated expansion that occurs in a very short period of time. Because of this expansion, the scenario could solve the problems of the hot big bang theory, such as the horizon and flatness problems. In addition, the scenario provides an explanation for the large-scale structure of the Universe by predicting the cosmological perturbations. \\ 
Although the scenario has been very successful in explaining and solving the problems, it suffers from problems such as the origin of the inflationary field, the problem of initial conditions. The scenario is also incomplete due to the big bang singularity and the collapse of classical general relativity at the Planck scale. The scenario is based on the classical description of spacetime and is assumed to occur at the energy scale of $10^3$ (or lower) orders of magnitude smaller than the Planck energy scale, where the initial condition for the scenario is also set at the beginning of inflation. But what happens in the pre-inflationary phase? This is an interesting question that cannot be answered by classical general relativity, and requires a quantum theory of gravity. \\  
Over the past decade, there has been increasing interest in loop quantum gravity (LQG) as a quantum theory of gravity. LQG is a non-perturbative, background independent approach to quantize gravity \cite{rovelli2004quantum,thiemann2008modern}. It imposes the fundamental discreteness at the Planck scale, so that the observable geometric quantities such as area and volume are discrete. loop quantum cosmology (LQC) is an application of LQG techniques to symmetry-reduced spacetime \cite{Bojowald:2008zzb,Giesel:2011idc}. The theory has been applied to various cosmological spacetimes and black holes (see \cite{Ashtekar:2011ni,Ashtekar:2018lag} for more details). The results are very encouraging. There is a generic resolution for the curvature singularity, where the singularity is replaced by a bounce, which is a consequence of the underlying discrete structure of quantum geometry \cite{Singh:2009mz,Singh:2014fsy}. Consequently, the universe undergoes a contracting phase before the bounce and an expanding phase after the bounce. \\ 
The dynamics is governed by a Hamiltonian constraint, which receives some modifications from LQC. The resulting Friedmann equation from the effective Hamiltonian contains a quadratic density modification as $H^2 \propto \rho (1 - \rho/\rho_c)$, which would be effective in the high energy regime \cite{Ashtekar:2006wn,Ashtekar:2007em,Singh:2012zc,Diener:2014mia,Zhu:2016dkn}. There is then a generic bounce as the energy density reaches the critical energy density denoted by $\rho_c = 0.41 m_p^4$. The framework has been widely used to study the pre-inflationary and inflationary phases using the canonical scalar field \cite{Ashtekar:2009mm,Agullo:2013ai,Agullo2023,Zhu:2017jew}. A numerical study of the pre-inflationary dynamics in loop quantum Brans-Dicke has also been investigated in \cite{Jin:2018wdx}.   \\

Besides the minimal scalar field, there are some other candidate fields containing non-standard kinetic terms, which are generally classified as k-essence. The properties of the tachyon field as a subclass of k-essence were first studied in LQC in \cite{Sen:2006uk}. It was then further explored in \cite{Xiong:2007ak,Xiao:2014nda,Xiao:2020olb}, where the authors found that the model could describe successful inflation with a sufficient number of a e-folds. They also showed that the quantum dynamical solution has attractive behavior. Another interesting subclass of k-essence models is known as the DBI scalar field. Although the DBI scalar field is known as a model inspired by string theory, here we follow a phenomenological perspective and are interested in investigating the cosmological consequences of the DBI scalar field in the framework of LQC \cite{Kinney:2007ag}. The dynamics of the DBI scalar field with exponential potential in LQC has been studied in \cite{Bhadra:2012qj}, where the field is considered as a dark energy model and the authors tried to account for the late-time behavior of the universe. However, there is no proper study of the DBI scalar field for the early times in the framework of LQC. More generally, the role of the k-essence in the LQC has not been properly studied. Motivated by this, we aim to fill this gap and provide a detailed investigation of the early universe evolution in the frame of LQC, where the DBI scalar field is the dominant field of the universe. Studies in standard cosmology indicates that the DBI scalar may be responsible for the very early universe evolution. Therefore, it would be interesting to consider such a more general field in a more general theory. The Lagrangian of the DBI field includes a non-standard kinetic term and a potential, plus it includes another function of the field, as $f(\phi)$, known as the wrap factor. Then there is some deviation in the Hamiltonian equations compared to the standard ones. In this work, the wrap function is taken as $f(\phi) = f_0 / \phi^4$, which is one of the most common choices in standard inflation. The potential of the field is also taken as a quadratic function of the scalar field, $V(\phi) = m^2 \phi^2 / 2$. It is found that there is a short period of super-inflation just after the bounce time, where the Hubble parameter increases and $\dot{H}$ is positive. After the time elapses, the universe enters the slow-roll inflationary phase, where the universe undergoes about $60$ e-fold expansion. Looking at the phase space trajectory of the field determines that the quantum solution contains attractive behavior.    \\ 

This paper aims to investigate the pre-inflationary dynamics in the frame of LQC, assuming that the DBI scalar field drives the universe in the early times. Section \ref{sec:dbi} briefly introduces the DBI scalar field, its Lagrangian, corresponding energy density and pressure, and obtains the momentum conjugate of the field necessary to generate the Hamiltonian. Section \ref{sec:dbiLQC} introduces the Hamiltonian equations. To obtain a numerical solution, the initial conditions are set at the bounce. Both the pre-inflationary and inflationary phases, amount of e-fold, and the field behavior are discussed. Section \ref{sec:phasespace} explains the phase space trajectory of the field, and the results are summarized in Section 
\ref{sec:conclusion}.  \\

\section{DBI scalar field}\label{sec:dbi}
The action is given by 
\begin{equation}
    S = \int d^4x \sqrt{-g} \; \mathcal{L} 
\end{equation}
and the Lagrangian for the DBI scalar field is read as \cite{Silverstein:2003hf,Copeland:2010jt}
\begin{equation}
	\mathcal{L} = {-1 \over f(\phi)} \; \sqrt{1 - f(\phi) \; \dot{\phi}^2 } \; +  {1 \over f(\phi)} - V(\phi)
\end{equation}
where $\phi$ is the DBI scalar field, and $\dot{\phi}$ is the time derivative of the field. $V(\phi)$ is the self-interacting potential and $f(\phi)$ is the wrapped brane tension function. The field contains a non-canonical kinetic term. It is assumed that the line element is described by a spatially flat (3+1) dimensional homogeneous and isotropic FLRW, read as
\begin{equation}
    ds^2 = -dt^2 + a^2(t) \Big( dx^2 + dy^2 + dz^2 \Big),
\end{equation}
in which $a^2(t)$ stands for the scale factor of the universe. Taking derivative with respect to the metric, one could obtain the corresponding energy density and pressure of the field as 
\begin{eqnarray}
    \rho_\phi & = & {\xi - 1 \over f(\phi)} + V(\phi) \\
    P_\phi & = & {\xi - 1 \over \xi \; f(\phi)} - V(\phi).
\end{eqnarray}
where the parameter $\xi$ is known as Lorentz factor and given by\footnote{The Lorentz factor usually is indicated by $\gamma$ in literature, however, here to avoid confusion with the Barbero-Immirzi parameter, which also shown by $\gamma$, we show the Lorentz factor by the parameter $\xi$.}
\begin{equation*}
    \xi = {1 \over \sqrt{1 - f(\phi) \; \dot{\phi}^2 }}
\end{equation*}

The momentum conjugate for the DBI field is calculated as
\begin{equation}
    \pi_\phi = {\partial \mathcal{L} \over \partial \dot{\phi}} 
              = {a^3 \; \dot\phi \over \sqrt{1 - f(\phi) \; \dot\phi^2}}
\end{equation}
Utilizing the Legendre transformation, the Hamiltonian for the DBI scalar field is read as
\begin{equation}
    \mathcal{H}_{DBI}= v \; \left( {\sqrt{v^2 + f(\phi) \; \pi_\phi^2} \over f(\phi) \; v} - {1 \over f(\phi)} + V(\phi) \right),
\end{equation}
where $v$ is the volume operator, $v = a^3$. We use this Hamiltonian in the next section to produce the Hamiltonian equations of the model. 
Taking the time derivative of the first equation, and using the second equation, the equation of motion of the DBI scalar field is obtained as
The equation of motion of the field is obtained by varying the action of the model with respect to the field, which is given by
\begin{equation}\label{field_eom}
    \ddot{\phi} + {3 H \over \xi^2} \; \dot{\phi} +  {3 f'(\phi) \over 2 f(\phi)} \; \dot{\phi}^2 - {f'(\phi) \over f^2(\phi)} + {1 \over \xi^3} \left( V(\phi) + {f'(\phi) \over f^2(\phi)} \right) = 0.
\end{equation} 
Besides the above usual method for deriving the energy density and pressure of DBI scalar field through the Lagrangian, one could obtain them from the Hamiltonian formalism as
\begin{eqnarray}
    \rho_\phi & = & {\mathcal{H}_{DBI} \over v}  = {\sqrt{v^2 + f(\phi) \; \pi_\phi^2} \over f(\phi) \; v} - {1 \over f(\phi)} + V(\phi) \\
    P_\phi & = & - {\partial \mathcal{H}_{DBI} \over \partial v} = {-v \over f(\phi) \; \sqrt{v^2 + f(\phi) \; \pi_\phi^2}}   + {1 \over f(\phi)} - V(\phi). \nonumber
\end{eqnarray}
This definition of the energy density and pressure of the DBI scalar field based on the Hamiltonian is helpful for the next section.

\section{Effective dynamics: DBI in loop quantum cosmology}\label{sec:dbiLQC}
The SU(2) Ashtekar-Barbero connection $A^i_a$ and the conjugate triad $E^a_i$ are the elementary classical phase space variables for the gravitational sector. Fixing the Gauss and diffeomorphism constraints, the only relevant constraint left is the Hamiltonian constraint, where we have assumed a homogeneous and isotropic universe. Vanishing the Hamiltonian constraint gives the physical solutions. The Hamiltonian of the gravitational sector has two parts, the Euclidean terms and the Lorentzian term. \cite{Ashtekar:2003hd,Ashtekar:2006wn}
\begin{equation}
    \mathcal{H}_{grav} = \mathcal{H}_{grav}^{(E)} - (1 + \gamma^2) \; \mathcal{H}_{grav}^{(L)}
\end{equation}
in which $\mathcal{H}_{grav}^{(E)}$ stands for the Euclidean term,
\begin{equation}
    \mathcal{H}_{grav}^{(E)} = \frac{1}{2} \int \mathrm{d}^3 x \, \epsilon_{ijk} F^i_{ab} \frac{E^{aj} E^{bk}}{\sqrt{\mathrm{det}(q)}}, 
\end{equation}
and 
\begin{equation}
    \mathcal{H}_{grav}^{(L)} = \int \mathrm{d}^3 x \,   K^j_{[a} K^k_{b]}   \frac{E^{aj} E^{bk}}{\sqrt{\mathrm{det}(q)}}, 
\end{equation}
represents the Lorentzian term (here it is assumed that the laps function is unity). 
$F^i_{ab}$ is the strength of the connection $A^i_a$, $K^i_a$ is the extrinsic curvature, and $|\mathrm{det}(q)|$ is the determinant of the space metric compatible with the triad. In LQC, the Lorentzian term is treated as a multiple of the Euclidean term. However, if one takes a different approach and treats the term differently, the modified version of LQC appears (see \cite{Ashtekar:2003hd,Ashtekar:2006wn,Li:2018opr,Li:2018fco,Li:2019ipm,Li:2019qzr,Li:2020mfi,Li:2019qzr,Li:2021mop,Li:2023dwy} for more details). \\
Including the DBI scalar field as the matter sector, the effective Hamiltonian is given by
\begin{equation}
    \mathcal{H} = {-3 \over 8 \pi G} \; {1 \over \gamma^2 \lambda^2} \; |\mathfrak{p}|^{1/2} \sin^2(\bar{\mu} \mathfrak{c}) + \mathcal{H}_{DBI}
\end{equation}
in which $\mathfrak{c}$ and $\mathfrak{p}$ are the respectively connection and triad variable obtained after the symmetry reduced. They satisfy $\{\mathfrak{c},\mathfrak{p}\}={8\pi G\gamma}/{3}$ where $\bar \mu\equiv \sqrt{\Delta/|\mathfrak{p}|}$ and $\Delta = \lambda^2 = 4\sqrt{3}\pi\gamma \ell_{\mathrm{Planck}}^2$ is the smallest non-zero eigenvalue of the area operator in LQG. It is found to be more convenient to work with the volume operator $v$ instead of the area operator $p$. Therefore, we go to a new set of phase space variable $b=\mathfrak{c}/\mathfrak{p}^{1/2}$ and $v=\mathfrak{p}^{3/2}=v_o a^3$ obtained by a canonical transformation. The effective Hamiltonian could be written as
\begin{equation}
    \mathcal{H} = -\frac{3v}{8 \pi G \lambda^2 \gamma^2} \; 
                      \sin^2\left(\lambda b\right) + \mathcal{H}_{DBI}
\end{equation}
The resulted Hamiltonian equations from the above Hamiltonian are acquired as
\begin{eqnarray}
\dot{v}  & = & \{ v, \mathcal{H} \} = {3 v \over 2 \gamma \lambda} \; \sin(2 \lambda \; b), \label{v_equation}   \\
\dot{b}  & = & \{ b, \mathcal{H} \} = {-3 \over 2 \gamma \lambda^2} \; \sin^2(\lambda \; b) - (4\pi G \gamma) P_\phi,  \label{b_equation}   \\
\dot{\phi}  & = & \{ \phi, \mathcal{H} \} = {\pi_\phi \over \sqrt{v^2 + f(\phi) \; \pi_\phi^2}},  \label{phi_equation} \\
\dot{\pi}_\phi & = & \{ \pi_\phi, \mathcal{H} \}  =  -v V'(\phi)  + \left( {v^2 + {1 \over 2} f(\phi) \; \pi_\phi^2 \over \sqrt{v^2 + f(\phi) \; \pi_\phi^2}} \right) {f'(\phi) \over f(\phi)}. \label{Pphi_equation} .
\end{eqnarray}
The Hubble parameter is given by $H = \dot{a} / a = \dot{v} / 3v$, which from Eq.\eqref{v_equation}, one could get the Friedmann equation as
\begin{equation}\label{Hubble2}
    H^2 = \left( {\dot{a} \over a} \right)^2 = \left( {\dot{v} \over 3 v} \right)^2 
         = {1 \over 4 \gamma^2 \lambda^2} \; \sin^2( 2 \lambda \; b)
\end{equation}
which are in terms of the phase space variable $v$ and $b$. Using the vanishing of the Hamiltonian constraint, one could get the energy density in terms of the phase space variable $v$ and $b$, which is
\begin{equation}
    \rho_\phi = \rho_c \; \sin^2(\lambda b)
\end{equation}
which $\rho_c$ stands for the critical density given by $\rho_c=3/(8\pi G \lambda^2\gamma^2)$. Working with the above equation, the phase space parameter $b$ is achieved in terms of the energy density, as 
\begin{equation}
    \sin^2(\lambda b) = {\rho_\phi \over \rho_c},
\end{equation}
and put it back to Eq.\eqref{Hubble2}, the Friedmann equation is rewritten as
\begin{equation}\label{Friedmann}
    H^2 = \frac{8\pi G }{3} \; \rho_\phi \; \left( 1 - \frac{\rho_\phi}{\rho_c}\right).
\end{equation}
Taking the time derivative of Eq.\eqref{v_equation}, using Eq.\eqref{b_equation} and the definition of the Hubble parameter, $H = \dot{v} / 3v$,  the time derivative of the Hubble parameter is achieved
\begin{equation}\label{dHdt}
    \dot{H} = - (4 \pi G) \; (\rho_\phi + P_\phi) \; \left( 1 - \frac{2 \rho_\phi}{\rho_c}\right),
\end{equation}
and the acceleration equation is obtained as
\begin{equation}\label{acceleration}
    {\ddot{a} \over a} = H^2 + \dot{H} = \frac{-4\pi G}{3} \; \rho_\phi \left( 1 - \frac{4 \; \rho_\phi}{\rho_c} \right)  - (4 \pi G) \; P_\phi\; \left( 1 - \frac{2 \; \rho_\phi}{\rho_c} \right)
\end{equation}
From Eqs.\eqref{Friedmann} and \eqref{acceleration}, the energy conservation relation is obtained which holds the same shape as in the standard cosmology
\begin{equation}\label{conservationlaw}
    \dot{\rho}_\phi + 3 H (\rho_\phi + P_\phi) = 0.
\end{equation}

\begin{figure}
\centering
\subfigure[]{\includegraphics[width=8.2cm]{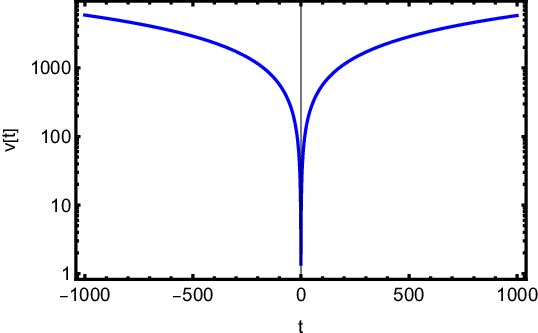}}
\subfigure[]{\includegraphics[width=8cm]{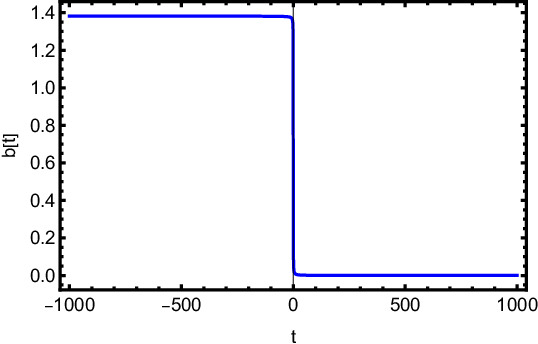}}
\subfigure[]{\includegraphics[width=8cm]{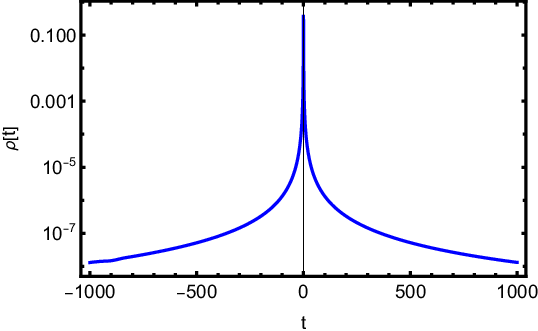}}
\caption{\label{vbbehavior}
The figure determines the behavior of the geometric parameters $v$ and $b$, and also the DBI scalar field energy density in both pre-bounce and post-bounce phases. The initial time is set at the bounce time $t_B = 0$, where the field is assumed to be $\phi_B = 1.51$. The other constants of the model are chosen as $f_0 = 1$ and $m = 1.21 \times 10^{-6}$. The first plot determines that the volume of the universe rapidly increases by time and consequently, the energy density rapidly decreases.  }
\end{figure}

We now concentrate on the numerical solution of the model. There is a four-dimensional phase space composed of $\{ v, b, \phi, \pi_\phi \}$, governed by a narrow set of first-order differential equations Eqs.\eqref{v_equation}, \eqref{b_equation}, \eqref{phi_equation} and \eqref{Pphi_equation}. To solve the equations one needs a set of initial equations for the parameter. We set the initial condition at the bounce time, so that we only need to specify the field at the bounce. Due to the scale symmetry of FLRW spacetime, one could scale the volume (or in other words, the scale factor) back to $v_B = 1$ at the bounce without loss of generality. At the bounce we also have $\dot{v} = 0$. On the other hand, the energy density at the bounce time is equal to the critical energy density $\rho_c$. Then, by determining the value of the field at the bounce, we have specified the potential, so one could determine the $\pi_\phi$ at the bounce time by solving the relation $\rho_c = \rho_\phi(t_B)$. Finally, we have the Hamiltonian constraint $\mathcal{H} = 0$, which should be satisfied in all time. This constraint gives us $b = \pi / 2 \lambda$ at the bounce time. By choosing the field $\phi(t_B) = 1.51 M_p$, the set of equations could be solved numerically, with the results shown in Fig.\ref{vbbehavior}. The behavior of the volume shows that the quantum bounce is symmetric, as expected, and the parameter $b$ is decreasing in all time. The energy density of the scalar field increases as it approaches the bounce point, and decreases after reaching the critical value $\rho_c$ at the bounce. \\

\begin{figure}
\centering
\subfigure[]{\includegraphics[width=8.2cm]{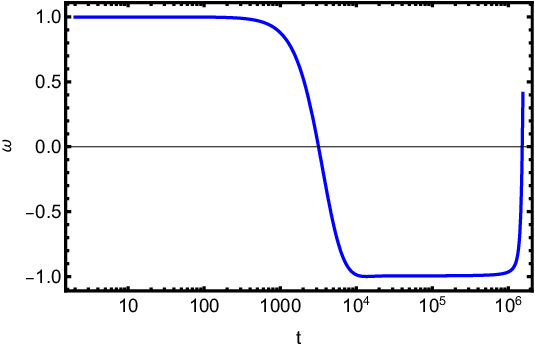}}
\subfigure[]{\includegraphics[width=8cm]{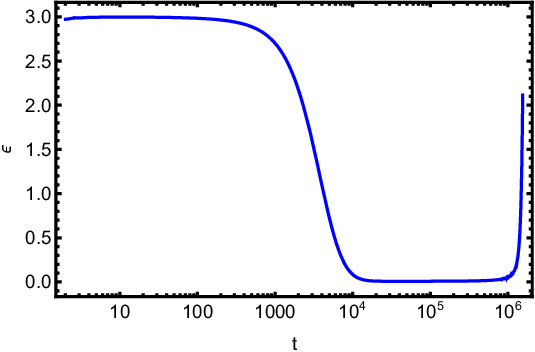}}
\subfigure[]{\includegraphics[width=8.4cm]{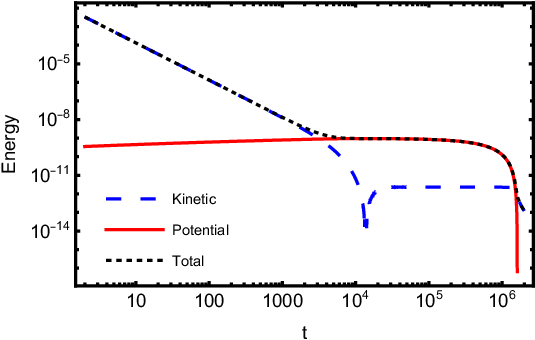}}
\caption{\label{omegaepsilon}
The behavior of the equation of state parameter of the DBI scalar field, the first slow-roll parameter, and the DBI field energy density is illustrated. The initial conditions are set the bounce time, same as the previous plot. The plot shows the dominant of the kinetic energy at the initial times, where $\omega = +1$ and $\epsilon >1$. Then, it decrease, where at the point $t = t_{eq}$, the kinetic energy and the potential energy become equal. Then, the potential dominates and the slow-roll inflationary phase starts. }
\end{figure}

The behavior of the equation of state parameter $\omega$ and the slow roll parameter $\epsilon$ is illustrated in Fig.\ref{omegaepsilon}. Observing the behavior of $\omega$, it can be seen that $\omega = 1$ near the bounce, indicating that the potential part of the field energy density is negligible. Therefore, from the beginning of the bounce to about $t = 1000 t_p$ (where $t_p$ is the Planck time), the kinetic part of the scalar field is dominant, and we call it the KE dominant region. Then, during a transition phase, the value of $\omega$ changes rapidly to $\omega = -1$. This indicates that the energy density of the field is dominated by the field. Then there are generally three phases: 
\begin{itemize}
    \item [i)] \textit{bounce phase:} The phase is from the bounce to the point where the quantum gravitational effect can be safely ignored. This is the point where the kinetic and potential parts of the field energy density are comparable, which in our case would be around $t = 1000 t_p$. From the energy density shown in Fig.\ref{omegaepsilon} one could see that the kinetic and the potential part are comparable and also the total energy density of the DBI scalar field is at the level where the quadratic term of the energy density in the Friedmann equation could be ignored and the equation could be well approximated by the classical Friedmann equation.

    \item[ii)] \textit{transition phase:} Then there is a transition phase, starting from the time when the equation of state parameter $\omega$ becomes zero until the start of slow-rolling inflation. The kinetic energy decreases during this phase and the potential energy begins to dominate. Compared to the other two phases, the transition phase is fast, so that the behavior of $\omega$ and $\epsilon$, as shown in Fig.\ref{omegaepsilon}, is similar to the step function. At the end of this phase, the kinetic energy of the field is negligible compared to the potential energy, and this would be the starting point of the slow inflationary phase. 

    \item[iii)] \textit{slow-roll inflation phase:} This phase begins when the slow-roll parameter $\epsilon$ reaches $\epsilon = 0.1$ for the first time after the transition phase\footnote{There is no precise definition for the start and end point of the slow-roll inflationary phase and one could find different choices in the literature, see \cite{Baumann:2009ds,Agullo:2015tca,Shahalam:2017wba,Li:2019ipm,Xiao:2020olb}. For example, \cite{Li:2019ipm} defines $|\eta_H| = 0.03$ as the starting point of inflation, whereas \cite{Xiao:2020olb} defines the starting point as the time when $\epsilon$ reaches $\epsilon = 0.1$.}. This accelerated expansion phase of the universe lasts until the slow-roll parameter $\epsilon$ reaches unity. $\epsilon = 1$ is a common definition for the end of the slow-roll inflationary phase, where the universe transitions from an accelerated expansion phase to a deceleration phase. For our case, from the behavior of $\epsilon$ in Fig.\ref{omegaepsilon}, it could be found that the inflation starts at the time around $t = 10^4 t_p$, where the potential energy of the field is much higher than the kinetic energy, shown in Fig.\ref{omegaepsilon}, and it lasts until the time around $t = 10^6 t_p$, where the slow-roll parameter $\epsilon$ reaches unity. 
\end{itemize}

\begin{figure}[h]
\centering
\subfigure[]{\includegraphics[width=8.2cm]{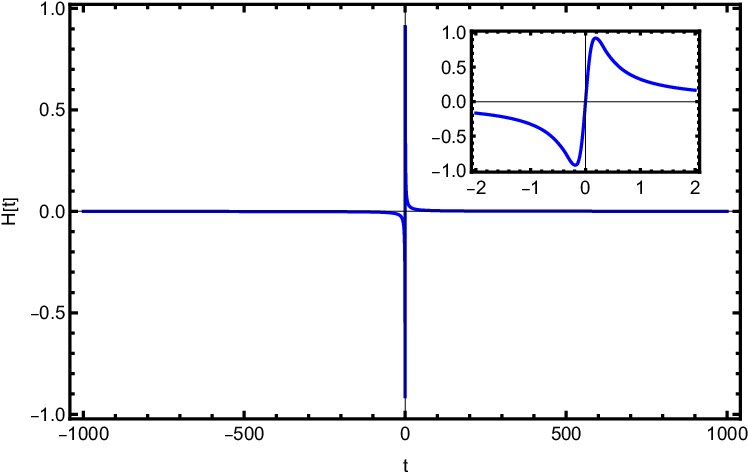}}
\subfigure[]{\includegraphics[width=8cm]{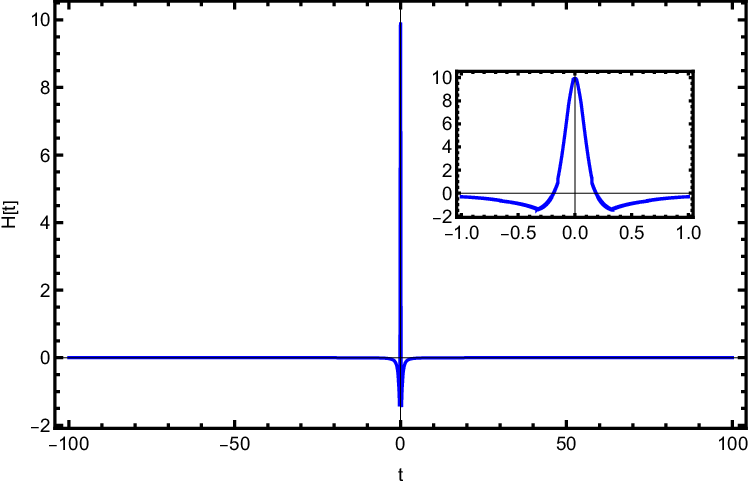}}
\caption{\label{Hubbledt}
The numerical evolution of the Hubble parameter and its first time derivative are plotted for the same initial conditions set at the bounce time. It is realized that the Hubble parameter starts to increase from the bounce time, and then reaches to a maximum value. These is the times where $\dot{H}$ is positive and known as the super-inflation. The inset plots are presented to clarify the behavior at the times close to the bounce. }
\end{figure}

\begin{figure}[h]
\centering
\includegraphics[width=8cm]{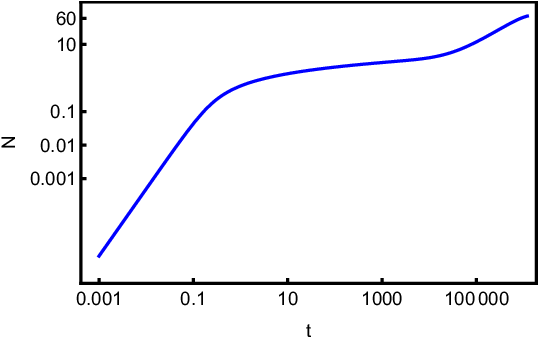}
\caption{\label{Nefold}
The plot shows the numerical result for the amount of the number of e-folds from the bounce time, where the initial conditions are the same as the previous plots; set the bounce }
\end{figure}

Considering the behavior of the Hubble parameter, we find that there is an interesting stage in the bounce phase where the time derivative of the Hubble parameter is positive, i.e. $\dot{H} > 0$. Fig.\ref{Hubbledt} indicates that there is a short period of time, starting as soon as the quantum bounce occurs, where not only the Hubble parameter but also $\dot{H}$ is positive. This stage in the evolution of the universe is known as super-inflation (SI). The SI stage also occurs in the case of the canonical scalar field and the tachyon scalar field considered in \cite{Xiao:2020olb}. This stage of expansion is known to be caused purely by quantum geometric effects. This stage of the SI ends when $\dot{H}$ reaches $\dot{H} = 0$. Exploring the number of e-folds in Fig.\ref{Nefold}, there are about $N = 0.11$ number of e-folds of expansion for the universe. From the start of the bounce to the transition time, the universe experiences about $N \simeq 4$ e-folds of expansion, and from the start of the slow-roll inflation to the end of this phase, we have about $N \simeq 59$ e-folds of expansion. Given that a successful inflationary model should provide about $55-65$ e-fold expansion, it is safe to say that the model can explain a desirable slow-roll inflation.  \\


\begin{figure}[h]
\centering
\includegraphics[width=8cm]{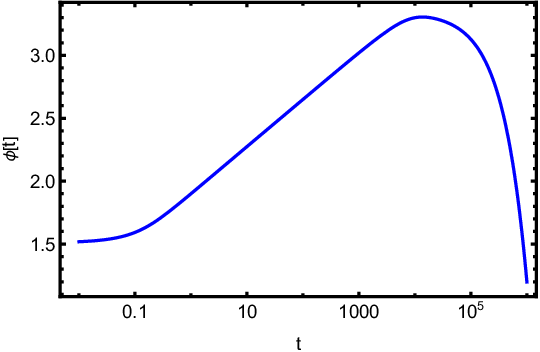}
\caption{\label{field}
The figure illustrates the numerical behavior of the DBI scalar field versus time for the post-bounce phase, with the same initial conditions set at the bounce, for $\dot{\phi}_B >0$. The field first start to increase meaning that the field is climbing up the potential. Next, it reaches to the maximum value, known as turnaround point. Then, it decreases, or in another word, it rolls down the potential.  }
\end{figure}

In the above solutions we assume a positive $\pi_\phi$ as the initial condition, but based on the Hamiltonian constraint one can also get the negative as the initial values for the parameter. Taking the positive sign for $\pi_\phi$ implies a positive $\dot\phi$, which indicates that the field will increase at the initial times after the bounce. As the field rises towards the top of the potential, $\dot\phi$ decreases and the potential increases. After some time, the kinetic energy and the potential energy reach the same value; in other words, the field reaches the turning point on the potential. This is the time when the slow roll inflation starts and the field slowly rolls down towards the minimum of the potential. Fig.\ref{field} illustrates the behavior of the field from the turning point to the end of the inflationary phase. The field initially takes up the potential, then reaches its maximum (the inflection point), then rolls down (note that the fast rolling shown in the table is only due to the scaling on the horizontal axes of the plot). \\

\begin{figure}[t]
\centering
\includegraphics[width=8cm]{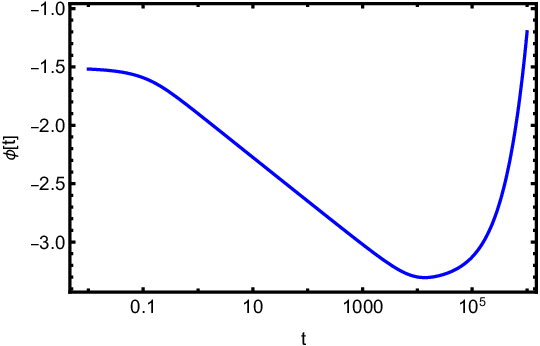}
\caption{\label{field_minus}
The figure illustrates the numerical behavior of the DBI scalar field versus time for the post-bounce phase, where the initial value of the field is chosen to be negative and $\dot{\phi}_B < 0$. The field starts to decrease  meaning that the field is climbing up the potential. Next, it reaches to the maximum value, known as turnaround point. Then, it increases, or in another word, it rolls down the potential.  }
\end{figure}

The Friedmann equations \eqref{Friedmann} and \eqref{dHdt}, as well as the field equation of motion \eqref{field_eom} are invariant under the transformation $(\phi, \dot{\phi}) \rightarrow (-\phi, -\dot{\phi})$. Solving the equations numerically with the initial conditions $\phi_B = -1.51 M_p$ and choosing the minus sign for $\pi_\phi$ at the bounce time leads to the same results for the volume $v(t)$, $b(t)$. However, we have a different case for the field. For this choice of initial conditions, the field is actually on the left wing of the potential at the bounce time. Since the initial $\dot\phi$ is also negative, the field decreases, i.e. it climbs up the potential. It then reaches the turning point where the potential energy is equal to the kinetic energy. At this point the slow roll inflation begins and the field slowly rolls down the potential. The behavior of the field is shown in Fig.\ref{field_minus}, where you can see that the field first decreases and then increases. \\

The results are summarized in Table.\ref{table_result}, where you could find the information for other choices of the field at the bounce time. It can be seen that for smaller values of the field at the bounce time, the SI phase lasts longer, but most importantly the slow-roll inflation phase lasts shorter, which leads to a lower number of e-folds. For example, for $\phi_B = 0.91$, the total number of e-folds is less than what is required for successful inflation. However, for $\phi_B = 1.51$, the universe experiences about $62$ e-folds of expansion during the slow-roll inflationary phase. \\

\begin{table}[h]
    \centering
 {\footnotesize
    \begin{tabular}{p{2.2cm}p{1.9cm}p{1.2cm}p{2.2cm}p{2.2cm}p{1.3cm}p{2.3cm}p{2cm}}
    \hline \\ [-0.15cm]
    Events & $t$  & $\phi$  & $\dot{\phi}$  & $H$  & $N$  & $V$  & $K$\\
    \hline \\ [-0.15cm]
    Bounce &  $0$   & $0.91$  & $0.6456$  & $1.13 \times 10^{-16}$  & $0$  & $6.06 \times 10^{-11}$  & $0.4093$ \\
    end of SI &  $0.20$   & $1.04$  & $0.5731$  & $0.9205$  & $0.1247$  & $7.88 \times 10^{-11}$  & $0.2118$ \\
    transition &  $4.30 \times 10^{3}$   & $2.63$  & $3.19 \times 10^{-5}$  & $9.72 \times 10^{-5}$  & $3.43$  & $5.07 \times 10^{-10}$  & $5.08 \times 10^{-10}$ \\
    onset of SRI &  $1.17 \times 10^{4}$   & $2.73$  & $4.29 \times 10^{-6}$  & $6.95 \times 10^{-5}$  & $4.006$  & $5.47 \times 10^{-10}$  & $9.23 \times 10^{-12}$ \\
    end of SRI &  $1.33 \times 10^{6}$   & $0.2096$  & $-1.73 \times 10^{-6}$  & $6.28 \times 10^{-6}$  & $52.56$  & $3.21 \times 10^{-12}$  & $1.50 \times 10^{-12}$ \\[0.2cm]
    \hline \\ [-0.15cm]
    Bounce &  $0$   & $1.15$  & $0.7749$  & $1.13 \times 10^{-16}$  & $0$  & $9.68 \times 10^{-11}$  & $0.4093$ \\
    end of SI &  $0.1899$   & $1.2867$  & $0.6089$  & $0.9203$  & $0.120$  & $1.21 \times 10^{-10}$  & $0.2075$ \\
    transition &  $3.87 \times 10^{3}$   & $2.86$  & $3.46 \times 10^{-4}$  & $1.09 \times 10^{-4}$  & $3.394$  & $5.99 \times 10^{-10}$  & $6.018 \times 10^{-10}$ \\
    onset of SRI &  $1.15 \times 10^{4}$   & $2.96$  & $3.53 \times 10^{-6}$  & $7.56 \times 10^{-5}$  & $4.036$  & $6.43 \times 10^{-10}$  & $6.24 \times 10^{-12}$ \\
    end of SRI &  $1.40 \times 10^{6}$   & $0.286$  & $-1.8 \times 10^{-6}$  & $7.98 \times 10^{-6}$  & $60.38$  & $5.98 \times 10^{-12}$  & $1.62 \times 10^{-12}$ \\ [0.2cm]
    \hline\\[-0.15cm]
    Bounce &  $0$   & $1.51$  & $0.8551$  & $1.13 \times 10^{-16}$  & $0$  & $1.68 \times 10^{-10}$  & $0.4093$ \\
    end of SI &  $0.1858$   & $1.6523$  & $0.6244$  & $0.9203$  & $0.119$  & $1.99 \times 10^{-10}$  & $0.2035$ \\
    transition &  $3.44 \times 10^{3}$   & $3.20$  & $3.88 \times 10^{-5}$  & $1.24 \times 10^{-4}$  & $3.350$  & $7.51 \times 10^{-10}$  & $7.53 \times 10^{-10}$ \\
    onset of SRI &  $1.05 \times 10^{4}$   & $3.31$  & $3.73 \times 10^{-6}$  & $8.45 \times 10^{-5}$  & $4.01$  & $8.02 \times 10^{-10}$  & $6.96 \times 10^{-12}$ \\
    end of SRI &  $1.13 \times 10^{6}$   & $1.13$  & $-1.9 \times 10^{-6}$  & $2.83 \times 10^{-5}$  & $66.06$  & $9.40 \times 10^{-11}$  & $1.90 \times 10^{-12}$ \\ [0.2cm]
    \hline
    \end{tabular}
}
    \caption{The table provide a brief results of the numerical solution for different values of the DBI scalar field set at the bounce time, where $\dot{\phi}$ is taken to be positive. Here, SI stands for super inflation stage and SRI implies on slow-roll inflation. The duration of the slow-roll inflationary phase and also the number of e-folds increase by enhancement of the field value at the bounce. These results are invariant under the transformation $(\phi, \dot{\phi}) \rightarrow (-\phi, -\dot{\phi})$. }
    \label{table_result}
\end{table}


\section{Evolution of the DBI scalar field}\label{sec:phasespace} 
\begin{figure}
\centering
\subfigure[\label{phaseQuadrant1}]{\includegraphics[width=8cm]{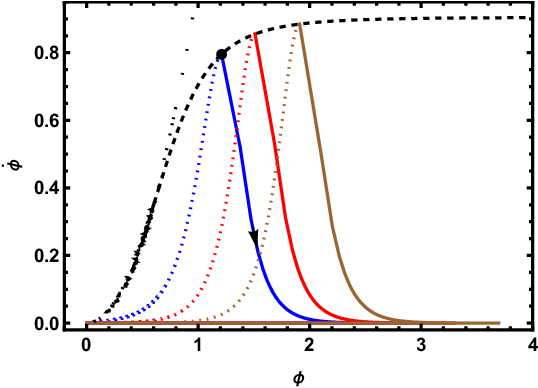}}
\subfigure[\label{phaseQuadrant3}]{\includegraphics[width=8cm]{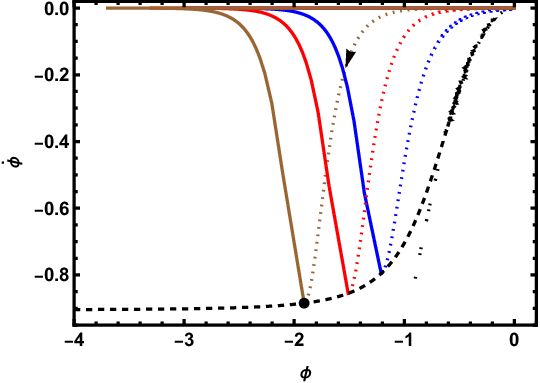}}
\caption{\label{phaseQuadrant}
The phase portrait of the DBI scalar field for three different values of the field with a) both $\phi$ and $\dot{\phi}_B$ are positive, and b) both $\phi$ and $\dot{\phi}_B$ are negative. The dashed black line indicates the bounce point, the solid color lines illustrate the phase space trajectories in the expanding phase and the dotted color lines indicates the phase space trajectories in the contracting phase.  }
\end{figure}

Here we will study the evolution of the DBI field in the context of LQC. Due to the complexity of the dynamical equation, it would be difficult to obtain an analytical solution, so we follow a numerical approach to gain more insight into this matter. The four independent variables of the phase space, i.e, 
$(v(t), b(t), \phi(t), \pi_\phi)$, are reduced to three by the Hamiltonian constraint. To consider the evolution of the DBI scalar field, the two-dimensional phase portrait $(\phi, \dot{\phi})$ will be studied. The bounce will be our critical point, so that the universe enters the expanding phase as time goes forward, and is in the contracting phase as time goes backward. The evolution of the field is determined by the DBI field equation of motion \eqref{field_eom}. Choosing the same initial condition as discussed in Sec.\ref{sec:dbiLQC}, the resulting two-dimensional phase portrait is shown in Fig.\ref{phaseQuadrant1}, where $\dot{\phi}_B >0$. The black dashed line is a set of points where the bounce occurs. It connects the contracting phase with the expanding phase of the universe. The black dot indicates the bounce point for each line, and the color lines illustrate the phase space trajectories for three different initial values of the DBI field. The solid lines are the trajectories in the expanding phase, and the arrow indicates the direction of increasing time. It can be seen that the trajectories approach the same point for all three initial field values. The dotted lines show the phase space trajectories in the contracting phase, so that with increasing time the trajectories approach the bounce point. Fig.\ref{phaseQuadrant3} shows the phase space trajectories so that one chooses the negative field values and $\dot{\phi}_B<0$ at the bounce for the initial condition values. It can be seen that there is the same behavior for the trajectories in both the expanding and contracting phases. \\

\section{Conclusion}\label{sec:conclusion} 
The pre-inflationary dynamics of the universe has been discussed in the context of LQC by introducing the DBI scalar field, which is the dominant component of the universe. The LQC is an application of the LQG to symmetry-reduced spacetime, which shows a significant deviation from classical geometry in the high energy regime, and reduces to classical general relativity in the low energy regime. The dynamics of the universe is determined by the effective Hamiltonian, which gives a set of coupled equations. The matter sector of the Hamiltonian is assumed to be given for the DBI scalar field model. \\ 
The scalar field model has a non-canonical kinetic term, which is a combination of the scalar field $\phi$ and its time derivative $\dot{\phi}$. The Hamiltonian of the field was determined and substituted into the effective Hamiltonian of the model. The Hamiltonian equations were then derived, which showed some deviations from the canonical scalar field model. By setting the initial conditions at the bounce point, the dynamical equations were solved numerically. \\
The solutions showed that there are three general phases after the bounce: the bounce phase, the transition phase and the slow inflationary phase. In addition, there is a very short period of super-inflation just after the bounce, where $\dot{H}$ is positive. During this super-inflation, the universe expands for about $0.11$ number of e-folds. Looking at the state parameter equation, it changes in the range $\omega \in [-1,1]$, so that in the bounce phase it is very close to $+1$, and then during the transition phase it rapidly decreases and reaches $-1$. This is the time when the slow roll inflation begins. In this context, the behavior of the first slow-roll parameter has also been considered, which shows a consistency with the behavior of $\omega$. At the same time as $\omega$ approaches $-1$, the slow roll parameter $\epsilon$ becomes very small. Considering the amount of expansion through the e-fold parameter, the universe expands by about $4$ e-folds during the bouncing phase, and for the slow-roll inflationary phase the model predicts about $59$ e-folds of expansion.  \\ 
The phase space trajectory of the DBI scalar field was considered in the last section. By choosing different values of the initial conditions for the field at the bounce, it was found that the field initially increases as the $\dot{\phi}$ decreases. After reaching the maximum point for the scalar field, the field starts to decrease. The phase space trajectories for all there initial conditions converge to the same point, indicating that the solution has an attractive behavior. The behavior of the field in phase space is in agreement with Fig.\ref{field}, where one could find the behavior of the field versus time. It shows that the field increases, and since $\dot{\phi} > 0$, this means that the scalar field climbs up the potential. Then it stops when $\dot{\phi}$ becomes zero, and from that point on the slow-roll inflationary phase starts and the field slowly rolls down the potential and the values of $\dot{\phi}$ are small and negative.


\begin{acknowledgments}
AM would like to thank Prof. Anzhong Wang for the opportunity to work in his research group. He also appreciates Dr. Bao-Fei Li and Dr. Yogesh for their helpful insights and discussions, as well as the members of the ITPC for their warm hospitality.
\end{acknowledgments}


\bibliography{RefBib}

\end{document}